\documentclass[prl,floats,aps,superscriptaddress,twocolumn,narrowtext]{revtex4}
\usepackage{amssymb}

\usepackage{amsmath}
\usepackage{exscale}
\usepackage[mathscr]{eucal}
\usepackage{epsfig}

\begin{document}

\title{Lorenz number in high $T_{c}$ superconductors: evidence for
bipolarons}
\author{K. K. Lee}
\affiliation{IRC in Superconductivity, Cavendish Laboratory,
University of Cambridge, Cambridge, CB3 0HE, United Kingdom}
\author{A. S. Alexandrov}
\affiliation{Department of Physics, Loughborough University,
Loughborough LE11 3TU, United Kingdom}
\author{W. Y. Liang}
\affiliation{IRC in Superconductivity, Cavendish Laboratory,
University of Cambridge, Cambridge, CB3 0HE, United Kingdom}

\begin{abstract}
Strong electron-phonon interaction in the cuprates has gathered support over
the last decade in a number of experiments. While phonons remain almost
unrenormalised, electrons are transformed into itinerent bipolarons and
thermally excited polarons when the electron-phonon interaction is strong.
We calculate the Lorenz number of the system to show that the
Wiedemann-Franz law breaks down because of the interference of polaron and
bipolaron contributions in the heat flow. The model fits numerically the
experimental Hall Lorenz number, which provides a direct evidence for
bipolarons in the cuprates.
\end{abstract}
\maketitle

\draft

The discovery of high-temperature superconductors \cite{mul,chu} has broken
constraints on the maximum $T_{c}$ predicted by the conventional theory of
low-temperature superconducting metals and alloys. Understanding the pairing
mechanism of carriers and the nature of the normal state in the cuprates and
other novel superconductors has been a challenging problem of the Condensed
Matter Physics. A number of theoretical models have been proposed,  which
rely on different non-phononic mechanisms of pairing (see, for example \cite
{and,kiv}). On the other hand, over the last decade, increasing evidence for
the electron-phonon interaction has been provided by isotope effect
measurements \cite{ZHAO}, infrared \cite{mic0,ita,TIM} and thermal
conductivity \cite{COHN}, neutron scattering \cite{ega}, and more recently
by ARPES \cite{LANZ,CHAI}.

To account for the high values of $T_{c}$ in the cuprates, one has to
consider electron-phonon (e-ph) interactions, which are larger than those
used in the intermediate coupling theory of superconductivity \cite{ELIA}.
Regardless of the adiabatic ratio, the Migdal-Eliashberg theory of
superconductivity and the Fermi-liquid have been shown to breakdown at the
e-ph coupling constant $\lambda \approx 1$ \cite{ALEX}. The many-electron
system collapses into the small (bi)polaron regime at $\lambda \gtrsim 1$
with well separated vibration and charge-carrier degrees of freedom.
Although it might have been thought that these carriers would have a mass
too large to be mobile, the inclusion of the on-site Coulomb repulsion and
the poor screening of the long-range e-ph interaction do lead to $mobile$
intersite bipolarons \cite{alekor,tru}. Above T$_{c}$ the Bose gas of these
bipolarons is non-degenerate and below $T_{c}$ their phase coherence sets in
and superfluidity of the doubly-charged $2e$ bosons can occur. In this
picture, the thermally excited single polarons co-exist with the Bose gas. \
\ 

There is much evidence for the crossover regime at $T^{\ast }$ and normal
state charge and spin gaps in the cuprates \cite{mic}. These energy gaps
could be understood as being half of the binding energy $\Delta $ and the
singlet-triplet gap of preformed bipolarons, respectively \cite{SANF}. Many
other experimental observations were explained using the bipolaron model 
\cite{aleedw}. These include the Hall ratio, the Hall angle, ab and c-axis
resistivities, magnetic susceptibility, and angle-resolved photoemission.
The bipolaron model provides parameter-free fits of critical temperatures,
upper critical fields, explains a remarkable increase of the quasiparticle
lifetime below $T_{c}$\cite{aleden}, and the symmetry of the order parameter 
\cite{alesym} in many cuprates. Further evidence for bipolarons comes from a
parameter-free estimate of the renormalized Fermi-energy $\epsilon _{F}$\cite
{aleF}, which yields a value well below $100meV.$ It is so small that
pairing is certainly $individual$ in most cuprates, i.e. the bipolaron size
is smaller than the inter-carrier distance. This is the case in a (quasi)
two-dimensional system, if 
\begin{equation}
\epsilon _{F}\lesssim 2\pi \Delta .
\end{equation}
The normal-state pseudogap, experimentally measured in many cuprates, was
found as large as $\Delta /2\gtrsim 50meV$ \cite{mic}, so that Eq.(1) is
well satisfied in underdoped and probably also in optimally doped cuprates.
One should notice that a coherence length in the charged Bose gas is $not$
the size of a boson. It depends on the interparticle distance and the
mean-free path, \cite{SANF}, and might be as large as in the BCS
superconductors. Hence, it is incorrect to apply the ratio of the coherence
length to the inter-carrier distance as a criterion of the BCS-Bose liquid
crossover. The criterion of real-space pairing is given by Eq.(1).

Direct evidence for the existence of  charged $2e$ Bose liquid in the normal
state cuprate materials is highly desirable. In 1993 Mott and Alexandrov\cite
{NEV} discussed the thermal conductivity $\kappa $; the contribution from
the carriers given by the Wiedemann-Franz ratio depends strongly on the
elementary charge as $\sim (e^{\ast })^{-2}$ and should be significantly
suppressed in the case of $e^{\ast }=2e$ compared with the Fermi-liquid
contribution. As a result, the Lorenz number, $L=\left( e/k_{B}\right)
^{2}\kappa _{e}/(T\sigma )$ differs significantly from the Sommerfeld value $%
L_{e}=\pi ^{2}/3$ of the standard Fermi-liquid theory, if carriers are
bipolarons. Here $\kappa _{e}$, $\sigma $, and $e$ are the electronic
thermal conductivity, the electrical conductivity, and the elementary
charge, respectively. Ref. \cite{NEV} predicted a very low Lorenz number $%
L_{b}$ for bipolarons, $L_{b}=6L_{e}/(4\pi ^{2})\approx 0.15L_{e}$, due to
the double charge of carriers, and also due to their nearly classical
distribution function above $T_{c}$.

Unfortunately, the extraction of the electron thermal conductivity has
proven difficult since both the electron term, $\kappa _{e}$ and the phonon
term, $\kappa _{ph}$ are comparable to each other in the cuprates. Some
experiments have attemped to get around this problem in a variety of methods 
\cite{SALA,TAKE,ZHANG}. In particular, Takenaka et al.\cite{TAKE} found that 
$\kappa _{e}$ is constant or weakly $T$-dependent in the normal state of $%
YBa_{2}Cu_{3}O_{6+x}$. This approximately $T$-independent $\kappa _{e}$
therefore implies the violation of the Wiedemann-Franz law (since the
resistivity is found to be a non-linear function of temperature) in the
underdoped region. The breakdown of the Wiedemann-Franz law has been seen
also in other cuprates \cite{HILL,ANDO}.

More recently a new way to determine the Lorenz number has been realised by
Zhang et al.\cite{zha}, based on the thermal Hall conductivity. The thermal
Hall effect allowed for an efficient way to separate the phonon heat current
even when it is dominant. As a result, the ``Hall'' Lorenz number, $%
L_{H}=\left( e/k_{B}\right) ^{2}\kappa _{xy}/(T\sigma _{xy})$, has been
directly measured in $YBa_{2}Cu_{3}O_{6.95}$ because transverse thermal $%
\kappa _{xy}$ and electrical $\sigma _{xy}$ conductivities involve only the
electrons. Remarkably, the measured value of $L_{xy}$ just above $T_{c}$ is
about the same as predicted by the bipolaron model, $L_{xy}\approx
0.15L_{e}. $ However, the experimental $L_{xy}$ showed a strong temperature
dependence, which violates the Wiedemann-Franz law. This experimental
observation is hard to explain in the framework of any Fermi-liquid model.

In this letter we propose a theory of the Lorenz number in the cuprates
explaining the experimental results by Zhang et al \cite{zha}. Our
particular interest lies on the conclusions that the Wiedemann-Franz law is
violated in the cuprates in the temperature range below the crossover
temperature $T^{\ast }$. Here we demonstrate that the Wiedemann-Franz law
breaks down because of the interference of polaron and bipolaron
contributions to the heat transport. When thermally excited polarons are
included, the bipolaron model explains the violation of the Wiedemann-Franz
law in the cuprates and the Hall Lorenz number as seen in the experiment.

Thermally excited phonons and (bi)polarons are well decoupled in the
strong-coupling regime of the electron-phonon interaction\cite{SANF}, so
that the conventional Boltzmann equation for renormalised carries can be
applied. We make use of the $\tau -$approximation\cite{ANSE} in an electric%
\textrm{\ }field $\mathbf{E=\nabla }\phi \mathbf{,}$ a temperature gradient $%
\mathbf{\nabla }T,$ and in a magnetic field $\mathbf{B\perp }$ $\mathbf{E,}$ 
$\mathbf{\nabla }T$. The bipolaron and single-polaron non-equilibrium
distributions are found as 
\begin{equation}
f(\mathbf{k})=f_{0}(E)+\tau \frac{\partial f_{0}}{\partial E}\mathbf{v}\cdot
\left\{ \mathbf{F}+\Theta \mathbf{n}\times \mathbf{F}\right\} ,  \label{4}
\end{equation}
where $\mathbf{v=}\partial E/\partial \mathbf{k,}$ $\mathbf{F}=(E-\mu )%
\mathbf{\nabla }T/T+\mathbf{\nabla }(\mu -2e\phi )$ and $f_{0}(E)=[y^{-1}{%
\exp (E/T)-1]}^{-1}$ for bipolarons with the energy $E=k^{2}/(2m_{b})$, and
the Hall angle $\Theta =\Theta _{b}=$ $2eB\tau _{b}/m_{b},$ and $\mathbf{F}%
=(E+\Delta /2-\mu /2)\mathbf{\nabla }T/T+\mathbf{\nabla }(\mu /2-e\phi )$
and $f_{0}(E)=\{y^{-1/2}{\exp [(E+\Delta /2)/T]+1\}}^{-1}$ , $%
E=k^{2}/(2m_{p})$ and $\Theta =\Theta _{p}=$ $eB\tau _{p}/m_{p\text{ }}$for
thermally excited polarons. Here $m_{b,p}$ are the bipolaron and polaron
masses of $two$-dimensional carriers, $y=\exp (\mu /T),$ $\mu $ is the
chemical potential, $\hbar =c=k_{B}=1$, and $\mathbf{n=B/}B$ is a unit
vector in the direction of the magnetic field. Eq.(2) is used to calculate
the electrical and thermal currents induced by the applied thermal and
potential gradients as

\begin{equation}
j_{\alpha }=a_{\alpha \beta }\mathbf{\nabla }_{\beta }(\mu -2e\phi
)+b_{\alpha \beta }\mathbf{\nabla }_{\beta }T,
\end{equation}
\begin{equation}
w_{\alpha }=c_{\alpha \beta }\mathbf{\nabla }_{\beta }(\mu -2e\phi
)+d_{\alpha \beta }\mathbf{\nabla }_{\beta }T.
\end{equation}
Eq.(3) defines the current with the polaronic conductivity $\sigma
_{p}=e^{2}\tau _{p}n_{p}/m_{p},$ where the kinetic coeffficents are given by 
\begin{eqnarray}
a_{xx} &=&a_{yy}=\frac{1}{2e}\sigma _{p}(1+4A), \\
a_{yx} &=&-a_{xy}=\frac{1}{2e}\sigma _{p}(\Theta _{p}+4A\Theta _{b}),  \notag
\\
b_{xx} &=&b_{yy}=\frac{\sigma _{p}}{e}[\Gamma _{p}+\frac{\Delta -\mu }{2T}+ 
\notag \\
&&2A(\Gamma _{b}-\mu /T)],  \notag \\
b_{yx} &=&-b_{xy}=\frac{\sigma _{p}}{e}[\Theta _{p}\left( \Gamma _{p}+\frac{%
\Delta -\mu }{2T}\right) +  \notag \\
&&2A\Theta _{b}(\Gamma _{b}-\mu /T)]  \notag
\end{eqnarray}
Eq.(4) defines the heat flow with the coefficients given by 
\begin{eqnarray}
c_{xx} &=&c_{yy}=\frac{\sigma _{p}}{2e^{2}}[T\Gamma _{p}+\Delta /2+e\phi + \\
&&2A(T\Gamma _{b}+2e\phi )],  \notag \\
c_{yx} &=&-c_{xy}=\frac{\sigma _{p}}{2e^{2}}[\Theta _{p}(T\Gamma _{p}+\Delta
/2+e\phi )+  \notag \\
&&2A\Theta _{b}(T\Gamma _{b}+2e\phi )],  \notag \\
d_{xx} &=&d_{yy}=\frac{\sigma _{p}}{e^{2}}\{T\gamma _{p}+\Gamma _{p}(\Delta
-\mu /2+e\phi )+  \notag \\
&&(\Delta /2+e\phi )\frac{\Delta -\mu }{2T}+  \notag \\
&&A[T\gamma _{b}+\Gamma _{b}(2e\phi -\mu )-2e\phi \mu /T]\},  \notag \\
d_{yx} &=&-d_{xy}=\frac{\sigma _{p}}{e^{2}}\{\Theta _{p}[T\gamma _{p}+\Gamma
_{p}(\Delta -\mu /2+e\phi )+  \notag \\
&&(\Delta /2+e\phi )\frac{\Delta -\mu }{2T}]+  \notag \\
&&A\Theta _{b}[T\gamma _{b}+\Gamma _{b}(2e\phi -\mu )-2e\phi \mu /T)]\} 
\notag
\end{eqnarray}
Here 
\begin{equation*}
\Gamma =\frac{\int_{0}^{\infty }dEE^{2}\partial f_{0}/\partial E}{%
T\int_{0}^{\infty }dEE\partial f_{0}/\partial E}=\frac{2\Phi (z,2,1)}{\Phi
(z,1,1)}
\end{equation*}
and 
\begin{equation*}
\gamma =\frac{\int_{0}^{\infty }dEE^{3}\partial f_{0}/\partial E}{%
T^{2}\int_{0}^{\infty }dEE\partial f_{0}/\partial E}=\frac{6\Phi (z,3,1)}{%
\Phi (z,1,1)}
\end{equation*}
are numerical coefficients, expressed in terms of the Lerch transcendent $%
\Phi (z,s,a)=\sum_{k=0}^{\infty }z^{k}/(a+k)^{s}$ with $z=y$ in $\Gamma _{b}$%
, $\gamma _{b}$ and $z=-y^{1/2}\exp [-\Delta /(2T)]$ in $\Gamma _{p}$, $%
\gamma _{p}$, and $A=m_{p}\tau _{b}n_{b}/(m_{b}\tau _{p}n_{p})$ is the ratio
of the bipolaron and polaron contributions to the transport, \ which
strongly depends on the temperature. For simplictiy we neglect the spin gap,
which is small in the optimally doped cuprates\cite{mic}. Then the bipolaron
singlet and triplet states are nearly degenerate, so that the bipolaron and
polaron densities are expressed as 
\begin{equation}
n_{b}=\frac{2m_{b}T}{\pi }|\ln (1-y)|,
\end{equation}
\begin{equation}
n_{p}=\frac{m_{p}T}{\pi }\ln \left[ 1+y^{1/2}\exp \left( -\frac{\Delta }{2T}%
\right) \right] .
\end{equation}
Using the kinetic coefficients Eqs. (5) and (6) we obtain 
\begin{eqnarray}
\rho  &=&\frac{1}{\sigma _{p}(1+4A)}, \\
R_{H} &=&\frac{1+4A\Theta _{b}/\Theta _{p}}{en_{p}(1+4A)^{2}}, \\
L &=&\frac{L_{p}+4AL_{b}}{1+4A}+\frac{A\left[ 2\Gamma _{p}-\Gamma
_{b}+\Delta /T\right] ^{2}}{(1+4A)^{2}}, \\
L_{H} &=&\frac{L_{p}+4AL_{b}\Theta _{b}/\Theta _{p}}{1+4A\Theta _{b}/\Theta
_{p}}+  \notag \\
&&\frac{A(4A+\Theta _{b}/\Theta _{p})\left[ 2\Gamma _{p}-\Gamma _{b}+\Delta
/T\right] ^{2}}{(1+4A)^{2}(1+4A\Theta _{b}/\Theta _{p})}
\end{eqnarray}
for the in-plane resistivity, the Hall ratio, the Lorenz number and the Hall
Lorenz number, respectively, where $L_{p}=(\gamma _{p}-\Gamma _{p}^{2})$ and 
$L_{b}=(\gamma _{b}-\Gamma _{b}^{2})/4$ are the polaron and bipolaron Lorenz
numbers. In the limit of a pure polaronic system (i.e., $A=0$) the Lorenz
numbers, Eqs.(11), (12) are $L=L_{H}=L_{p}.$ In the opposite limit of a pure
bipolaronic system (i.e. $A=\infty $) we obtain a reduced Lorenz number\cite
{NEV} $L=L_{H}=L_{b}.$ However, in general our equations (11) and (12) yield
the $temperature$ $dependent$ Lorenz numbers that differ significantly from
both limits. The main difference originates in the second terms in the right
hand side of Eqs.(11) and (12), which describe an interference of polaron
and bipolaron contributions in the heat flow. In the low-temperature regime, 
$T\ll \Delta $, this contribution is exponentially small because the number
of unpaired polarons is small. However, it is enhanced by the factor $%
(\Delta /T)^{2},$ and becomes important in the intermediate temperature
range $T_{c}<T<T^{\ast }$. The contribution appears as a result of the
recombination of a pair of polarons into the bipolaronic bound state at the
cold end of the sample, which is reminicent to the contribution of the
electron-hole pairs to the heat flow in semiconductors \cite{ANSE}. These
terms are mainly responsible for the breakdown of the Wiedemann-Franz law in
the bipolaronic system.

It has been shown that the bipolaron model fits nicely the temperature
dependencies of the in-plane \cite{in} and out-of-plane\cite{out}
resistivities and the Hall ratio in the cuprates. Here we show that it also
fits the Hall Lorenz number measured by Zhang et al.\cite{zha}. To reduce
the number of fitting parameters we take the charge pseudogap $\Delta
/2=600K $, as found by Mihailovic et al.\cite{mic} for nearly optimally
doped $YBa_{2}Cu_{3}O_{6+x}$ in their systematic analysis of charge and spin
spectroscopies. According to Ref. \cite{BRAT} the main scattering channel
above $T_{c}$ is due to the particle-particle collisions with the relaxation
time $\tau _{b,p}\propto 1/T^{2}$. The chemical potential is pinned near the
mobility edge, so that $y\approx 0.6$ in a wide temperature range, if the
number of localised states in the random potential is about the same as the
number of bipolarons\cite{BRAT}. This is the case in $YBa_{2}Cu_{3}O_{6+x}$,
where every excess oxygen ion $x$ can localise the bipolaron. As a result,
there is only one fitting parameter in $L_{H}$, Eq.(12) which is the ratio
of the bipolaron and polaron Hall angles $\Theta _{b}/\Theta _{p}$. The
model well fits the experiment, Fig.1, with a reasonable value of $\Theta
_{b}/\Theta _{p}=0.44$. It also quantitatively reproduces the (quasi)linear
in-plane resistivity and the inverse Hall ratio, as observed in the cuprates
(upper inset in Fig.1 and Ref. \cite{BRAT,in}).

\begin{figure}[h]
\centering \epsfig{file=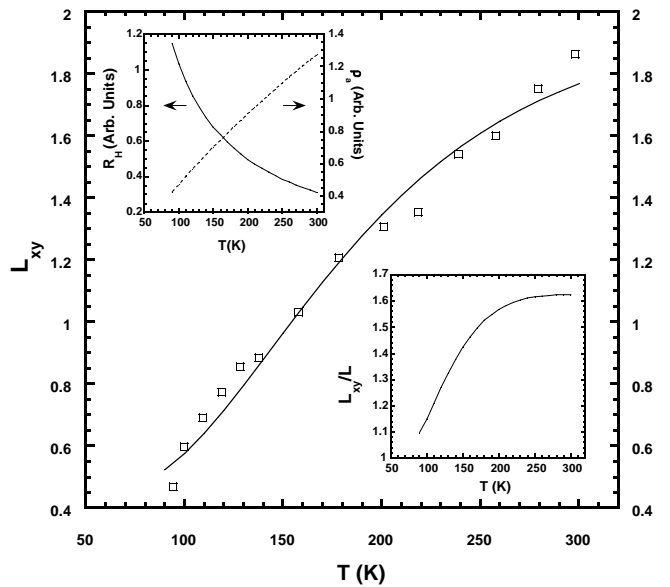,bbllx=28,
bblly=149, bburx=568, bbury=637,width=3.5in}
\caption{The experimental Hall Lorenz number \cite{zha} in $%
YBa_{2}Cu_{3}O_{6.95}$ fitted by the bipolaron theory, Eqs.(11,12),
with $\Theta _{b}/\Theta _{p}=0.44$. The upper inset shows the
linear in-plane resistivity and the Hall ratio. The lower inset
shows the ratio of the Hall Lorenz number to the Lorenz number as
a function of temperature.} \label{fig.1}
\end{figure}

Lower inset in Fig.1 shows a slightly lower Lorenz number compared with the
Hall Lorenz number produced by our calculations. Because the thermal Hall
conductivity directly measures the Lorenz number in the framework of
our model, it can be used to measure
the lattice contribution to the heat flow as well. When we subtract the
electronic contribution determined by using the Lorenz number, the lattice
contribution to the diagonal heat flow appears to be much higher than it is
anticipated in the framework of any Fermi-liquid model.

We notice that some recent measurements \cite{PROU} on $Tl_{2}Ba_{2}Cu0_{6+%
\delta }$ suggest that the Wiedemann-Franz law holds perfectly well in the
overdoped region and therefore conclude that the Fermi-liquid prevails at
this doping range. Alexandrov and Mott \cite{SANF}, suggested that there
might be a crossover from the Bose-Einstein condensation to a BCS-like $%
polaronic$ superconductivity across the phase diagram. Thus Proust et al's
results are still compatible with the (bi)polaron picture. If the Fermi
liquid does exist at overdoping then it is likely that the heavy doping
causes an ''overcrowding effect'' where the polarons find it difficult to
form bipolarons due to the larger number of competing holes\cite{SANF}.

We conclude that by the necessary inclusion of thermally excited polarons as
the temperature rises, the bipolaron model predicts the Lorenz number very
close to experiment in underdoped and optimally doped cuprates. Our
consideration leads to good fits for the experimental Hall Lorenz number,
Hall ratio, and the in-plane resistivity. The interference of the polaron
and bipolaron contributions to the energy flow breaks down the
Wiedemann-Franz law and results in the unusual temperature dependence of the
Lorenz number.

This work was supported by the Leverhulme Trust (grant F/00261/H) and by the
EPSRC UK (grant R46977). We greatly acknowledge P.W. Anderson and Y. Zhang
for helpful discussions of some theoretical and experimental aspects.


\begin{references}
\bibitem{mul}  J.G. Bednorz and  K.A. M\"{u}ller, Z. Phys. B{\bf 64}, 189 (1986)

\bibitem{chu}  M.K. Wu, J.R.  Ashburn,  C.J. Torng, P.H.  Hor, R.L. Meng, L. Gao,
Z.J. Huang,  Y.Q. Wang, and C.W. Chu, Phys. Rev. Lett. {\bf 58 } 908 (1987).

\bibitem{and}  P.W. Anderson, {\it The Theory of Superconductivity in the
High T$_c$ Cuprates}, (Princeton University Press, Princeton, 1997).

\bibitem{kiv}  E.W. Carlson, V.J. Emery,S.A.  Kivelson, and D. Orgad, {\it \
Concepts in High Temperature Superconductivity} cond-mat/0206217 and
references therein.

\bibitem{ZHAO}  G. Zhao, M.B. Hunt, H. Keller and K.A. Muller, Nature, {\bf %
385}, 236 (1997).

\bibitem{mic0}  D. Mihailovic, C.M. Foster, K. Voss and A.J. Heeger, Phys.
Rev. B{\bf 42}, 7989 (1990).

\bibitem{ita}  P. Calvani, M.Capizzi, S. Lupi, P. Maselli, A. Paolone, P.
Roy, S-W Cheong, W. Sadowski and E. Walker, Solid State Commun. {\bf 91},
113 (1994).

\bibitem{TIM}  T. Timusk, C.C. Homes and W. Reichardt, {\it \ Anharmonic
properties of High T$_c$ Cuprates}, eds. D.Mihailovic, G. Ruani, E. Kaldis
and K.A. Muller, 171(World Scientific, Singapore, 1995).

\bibitem{COHN}  J.L. Cohn, S. Wolf and T.A. Vanderah, Phys. Rev. B , {\bf 45}%
, 511 (1992).

\bibitem{ega}  T. Egami, J. Low Temp. Phys. {\bf 105}, 791(1996).

\bibitem{LANZ}  A. Lanzara, P.V. Bogdanov, X.J. Zhou, S.A. Kellar, D.L.
Feng, E.D. Lu, T. Yoshida, H. Eisaki, A. Fujimori,K. Kishio, J.I. Shimoyana,
T. Noda, S. Uchida, Z. Hussain and Z.X. Shen, Nature, {\bf 412}, 510 (2001).

\bibitem{CHAI}  A. Chainani, T. Yokoya, T. Kiss, S. Shin, T. Nishio and H.
Uwe, Phys. Rev. B, {\bf 64}, 180509 (2001).

\bibitem{ELIA}  G.M. Eliashberg, Sov.Phys. JETP, {\bf 11}, 696 (1960).

\bibitem{ALEX}  A.S. Alexandrov, Phys. Rev. B, {\bf 46}, 2838 (1992).

\bibitem{alekor} A.S.  Alexandrov  and P.E. Kornilovich, Phys. Rev. Lett.
{\bf 82,} 807 (1999); J. Phys.: Condens. Matter {\bf 14}, 5337 (2002).

\bibitem{tru}  J. Bonca J and S.A. Trugman, Phys. Rev. B {\bf 64}, 094507
(2001).

\bibitem{mic}  D. Mihailovic, V.V. Kabanov, K. Zagar, and J. Demsar, Phys.
Rev. B{\bf 60}, 6995 (1999) and references therein.

\bibitem{SANF}  A.S. Alexandrov and N.F. Mott, {\it High Temperature
Superconductors and Other Superfluids}, (Taylor and Francis, London, 1994).

\bibitem{aleedw}  for a recent review see A.S. Alexandrov and P.P. Edwards,
Physica C{\bf 331}, 97 (2000), and references therein.

\bibitem{BRAT}  A.S. Alexandrov, A.M. Bratkovsky and N.F. Mott, Phys. Rev.
Lett, {\bf 72}, 1734 (1994).

\bibitem{aleden}  A.S. Alexandrov and C. J. Dent, J. Phys.: Condens. Matter
{\bf 13}, pp L417 (2001).

\bibitem{alesym}  A.S. Alexandrov, Phil. Mag. B{\bf 81}, 1397 (2001).

\bibitem{aleF}  A.S. Alexandrov, Physica C {\bf 363}, 231 (2001).

\bibitem{NEV}  A.S. Alexandrov and N.F. Mott, Phys. Rev. Lett, {\bf 71},
1075 (1993).

\bibitem{TAKE}  K. Takenaka, Y. Fukuzumi, K. Mizuhashi, S. Uchida, H. Asaoka
and H. Takei, Phys. Rev. B, {\bf 56}, 5654 (1997).

\bibitem{SALA}  R.C. Yu, M.B. Salamon, J.P. Lu and W.C. Lee, Phys. Rev.
Lett., {\bf 69}, 1431 (1992).

\bibitem{ZHANG}  Y. Zhang, N.P. Ong, Z.A. Zhang, R. Gagnon and L. Taillefer,
Cond-mat/0001037 (2000).

\bibitem{HILL}  R.W. Hill, C. Proust, L. Taillefer, P. Fournier and R.L.
Greene, Nature, {\bf 414}, 711 (2001).

\bibitem{ANDO}  J. Takeya, Y. Ando, S. Komiya and X.F. Sun, Cond-mat/0108055
(2002).

\bibitem{zha}  Y. Zhang, N.P. Ong, Z.A. Xu, K. Krishana, R. Gagnon, and L.
Taillefer, Phys. Rev. Lett., {\bf 84}, 2219 (2000).

\bibitem{ANSE}  A. Anselm, {\it Introduction of Semiconductor Theory} ,
(Prentice and Hall, New Jersey, 1981).

\bibitem{in}  X.H. Chen, M.  Yu, K.Q. Ruan, S.Y. Li, Z. Gui,  G.C. Zhang, L.Z. Cao,  Phys.
Rev. B {\bf 58,} 14219 (1998);

W.M. Chen, J.P. Franck, and J. Jung, Physica C {\bf 341}, 1875 (2000).

\bibitem{out}  J. Hofer, J. Karpinski,  M. Willemin, G.I. Meijer, E.M. Kopnin, 
R. Molinski,  H. Schwer,  C. Rossel, and H.  Keller, Physica C {\bf 297}, 103 (1998);

V.N. Zverev and D.V. Shovkun, JETP Lett. {\bf 72}, 73 (2000).

\bibitem{PROU}  C. Proust, E. Boakin, R.W. Hill, L. Taillefer and A.P.
Mackenzie, Cond-Mat/0202101 (2002).
\end{references}
\end{document}